\documentclass[10pt,twocolumn]{autart}    

\usepackage[colorlinks,linkcolor=blue,anchorcolor=black,linktocpage=true,urlcolor=black,citecolor=black]{hyperref}  
\usepackage{natbib}
\usepackage{amsmath,amsfonts}
\usepackage{amssymb}
\usepackage{mathtools} 
\usepackage{graphicx}
\usepackage{textcomp}
\usepackage{xcolor}
\usepackage{soul}
\usepackage{url}
\usepackage{multirow}
\usepackage{colortbl}
\usepackage{float}
\usepackage{array}
\usepackage{wrapfig}

\graphicspath{{figures/}}
\newtheorem{theorem}{\bf Theorem}
\newtheorem{lemma}{\bf Lemma}
\newtheorem{remark}{\bf Remark}
\newtheorem{assumption}{\bf Assumption}
\allowdisplaybreaks 

\begin{document}

\begin{frontmatter}
\title{Adaptive Safety with Control Barrier Functions and Triggered Batch Least-Squares Identifier} % Title, preferably not more than 10 words.

\thanks[footnoteinfo]{This work was supported by National Natural Science Foundation of China under grants 62373019, 62022008 and 61973017. The material in this paper was not presented at any conference.}

\author[BUAA]{Jiajun Shen}\ead{jjshen@buaa.edu.cn},   
\author[BUAA,Zhongguancun]{Wei Wang\corauthref{cor1}}\ead{w.wang@buaa.edu.cn},               
\corauth[cor1]{Corresponding author.}
\author[Agder]{Jing Zhou}\ead{jing.zhou@uia.no}, 
\author[BUAA,Zhongguancun]{Jinhu  Lu}\ead{jhlu@iss.ac.cn} 

\address[BUAA]{School of Automation Science and Electrical Engineering, Beihang University, Beijing 100191, China}  % Please supply                                              
\address[Zhongguancun]{Zhongguancun Laboratory, Beijing 100194, China}  % Please supply 
\address[Agder]{Department of Engineering Sciences, University of Agder, Grimstad, Norway 4898, Norway}  % Please supply   

\begin{abstract}                          % Abstract of not more than 200 words.
In this paper, a triggered Batch Least-Squares Identifier (BaLSI) based adaptive safety control scheme is proposed for uncertain systems with potentially conflicting control objectives and safety constraints.
A relaxation term is added to the Quadratic Programs (QP) combining the transformed Control Lyapunov Functions (CLFs) and Control Barrier Functions (CBFs), to mediate the potential conflict.
The existing Lyapunov-based adaptive laws designed to guarantee specific properties of the Lyapunov functions, may grow unboundedly under the effects of the relaxation term.
The proposed adaptive law is designed by processing system inputs and outputs, to avoid unbounded estimates and overparameterization problems in the existing results.
A safety-triggered condition is presented, based on which  the forward invariant property of the safe set is shown and Zeno behavior can be excluded.
Simulation results are presented to demonstrate the effectiveness of the proposed adaptive control scheme.
\end{abstract}
\end{frontmatter}

\section{Introduction}
With the increasing applications of unmanned systems, frequent accidents have attracted the attention of researchers.
System security is generally guaranteed by constraining the states within a safe set.
Several transformed barrier functions, which tend to infinity at the boundary of the safe set, have been proposed to ensure the system security, such as Prescribed Performance Control  \citep{bechlioulis2008robust,wang2010adaptive} and Barrier Lyapunov Function  \citep{tee2009barrier}.
However, these control schemes are not suitable for the systems under potential conflict between control objectives and safety constraints.
Quadratic Program (QP) based control schemes are proposed to mediate this conflict, where the  control objectives and safety constraints are transformed to Control Lyapunov Functions (CLFs) \citep{sontag1989universal} and Control Barrier Functions (CBFs) \citep{ames2014control,ames2017control,xu2015robustness}.

This paper focuses on uncertain systems with potentially conflicting control objectives and safety constraints.
Much efforts have been devoted to develop robust and adaptive schemes for CLFs and CBFs.
The robust Control Lyapunov Functions (rCLFs) and robust Control Barrier Functions (rCBFs) are combined by QP in \citet{cohen2022robust}, with the assumption that the range of the uncertainties is known a priori.
The control signal must ensure that the necessary properties of rCLFs and rCBFs are met, irrespective of the values of the unknown parameters within the known range.
Thus these robust control schemes are also called ``worst-case uncertainty'' schemes.
The prior knowledge assumption and the conservative control performance are significant disadvantages of the robust schemes for QP-based controllers.
Several parameter update laws are proposed to reduce the effects of model uncertainties for the adaptive Control Lyapunov Functions (aCLFs) \citep{krstic1995control} and adaptive Control Barrier Functions (aCBFs) \citep{lopez2020robust,isaly2021adaptive,cohen2022high,wang2024adaptive,lopez2023unmatched}.
Since CBFs do not have the favourable positive definiteness property as Lyapunov functions, it is difficult to guarantee the boundedness of the estimates by employing Lyapunov-based adaptive schemes.
The existing results on aCBFs always robustly enforce the CBF condition using certain pre-known closed convex set of the unknown parameters, which leads their control performance to be highly conservative, as similar to the ``worst-case uncertainty'' schemes.
Lyapunov-based adaptive controllers are proposed in \citet{taylor2020adaptive}, which combines the aCLFs and aCBFs by QP.
However, the relaxation term $\delta$, which aims to mediate the potential conflict between control objectives and safety constraints, may cause the parameter estimate to grow unboundedly.
Moreover, the parameter estimates in \citet{taylor2020adaptive} are overparametrized, i.e. the number of parameter estimates is greater than the number of unknown parameters \citep{krstic1992adaptive}.

In \citet{krstic1995nonlinear}, the adaptive control schemes are classified into ``Lyapunov-based'' \citep{krstic1992adaptive,krstic1995control,taylor2020adaptive,isaly2021adaptive,lopez2020robust,cohen2022high} and ``estimation-based'' 
\citep{ioannou1996robust,cohen2023modular}.
The Lyapunov-based adaptive laws are designed to guarantee some desired properties of the Lyapunov functions.
In \citet{taylor2020adaptive}, the boundedness of Lyapunov functions cannot be guaranteed when the relaxation term $\delta$ breaks these properties, further the parameter estimates are not guaranteed to be bounded either.
In \citet{cohen2023modular}, an estimation-based parameter update law is proposed to address the problems of unboundedness and overparametrization estimates.
However, it establishes safety using an ISS framework, indicating that the safety constraints cannot be totally guaranteed subject to parameter estimation error.

A novel regulation-triggered Batch Least-Squares Identifier (BaLSI) is proposed in \citet{karafyllis2018adaptive,karafyllis2020adaptive}, whose adaptive laws are activated at the moment the control performance decreases.
Several event-triggering rules \citep{seyboth2013event,wang2020adaptive,wang2021adaptive}
are proposed to exclude the Zeno behavior, which indicates that infinite times of triggering in a finite time interval can be avoided.
Inspired by the BaLSI, this paper proposes a novel adaptive control scheme for the QP-based controllers with aCLFs and aCBFs.
Our contributions are three-folds.
\begin{itemize}  
	\item [$\bullet$] 
	Different from related results on aCLFs and aCBFs \citep{taylor2020adaptive,lopez2020robust}, the parameter estimate is derived by processing the data of system inputs and outputs.
	Thus the boundedness of  parameter estimate can be ensured, no matter how the relaxation term $\delta$ affects the system.
	\item [$\bullet$] 
	Different from Lyapunov-based adaptive schemes \citep{taylor2020adaptive,lopez2020robust,isaly2021adaptive,cohen2022high,wang2024adaptive,lopez2023unmatched}, the proposed adaptive update laws are designed to directly estimate the unknown parameters, rather than to guarantee the desired properties of CLFs and CBFs.
	The number of parameter estimates is always equal to the number of unknown parameters, hence the overparametrization problem can be solved.
	Moreover, the proposed scheme never enforce the CBF condition using pre-known closed convex set of the unknown parameters, thus it will not perform conservative performance like previous results.
	\item [$\bullet$] 
	Note that the parameter estimate is designed to be updated when the system states tend to the boundary of the safe set.
	The updating times of parameter estimate is less than the dimension of the unknown parameter vector, and the effects due to unknown system parameters will be totally eliminated after the last updating moment of the parameter estimate.
	A safety-triggered condition is presented, based on which  the forward invariant property of the safe set is shown and the Zeno behavior can be excluded.
\end{itemize}

The rest of this paper is organized as follows. 
In Section 2, we introduce some basic preliminaries and formulate the problems. 
In Section 3, the safety-triggered BaLSI-based adaptive control scheme is presented.
In Section 4, the system stability and some special properties of the proposed scheme are analyzed.
In Section 5, simulation results are provided.
A conclusion is drawn in Section 6.

\textbf{Notations.}
The sets of real numbers and non-negative real numbers are denoted by $\Re$ and $\Re^+$, respectively.
The set of positive integer numbers is denoted by $Z^+$.
Given a vector $x \in \Re^n$, $\|x\|$ denotes the Euclidean norm of $x$.
Given a matrix $G$, $\mathcal{R}[G]$ and $\mathcal{N}[G]$ denote the range space and null space of $G$ respectively, $\dim \mathcal{N}[G]$ denotes the dimension of $\mathcal{N}[G]$.
The null set is denoted by $\emptyset$.
$\operatorname{Int}(\mathcal{C})$ and $\partial \mathcal{C}$ denote the interior and boundary of the set $\mathcal{C}$, respectively.
A continuous function $\alpha : \left[0,a\right) \rightarrow \left[0, \infty\right)$ belongs to class $\mathcal{K}$ if it is strictly increasing and $\alpha(0)=0$. 
It belongs to class $\mathcal{K}_{\infty}$ if $a=\infty$ and $\alpha(r) \rightarrow \infty$ as $r \rightarrow \infty$.
A continuous function $\beta: (-b,a) \rightarrow (-\infty,\infty)$ belongs to extended class $\mathcal{K}$ for some $a,b \in \Re^+$ if it is strictly increasing and $\beta(0)=0$.

\section{Preliminaries and Problem Formulation}
Consider a dynamic system, which can be represented by the following model,
\begin{equation}
	\dot{x}=f(x)+\phi(x)^T\theta+g(x)u,
	\label{model}
\end{equation}
where $x \in \mathcal{X}$ and $u \in \Re^m$ are the state and control input, respectively.
$\mathcal{X} \subset \Re^n$ is the state space, which is assumed to be path-connected and $0 \in \mathcal{X}$.
$f: \mathcal{X} \rightarrow \Re^n$, $\phi: \mathcal{X} \rightarrow \Re^{p \times n}$ and $g: \mathcal{X} \rightarrow \Re^{n \times m}$ are known locally Lipschitz functions.
$\theta \in \Re^p$ denotes the vector of unknown constant parameters.

\subsection{Preliminaries}
The control objectives and safety constraints will be transformed to the requirements on Control Lyapunov Functions (CLFs) and Control Barrier Functions (CBFs), respectively.
Then the control input $u$ can be calculated by Quadratic Programs (QP). 
The CLFs and CBFs are defined below.\\
\\
\text{\bf Definition 1 \citep{sontag1995characterizations}.}
\textit{A continuously differentiable function $V: \mathcal{X} \rightarrow \Re^+$ is a Control Lyapunov Function if there exist class $\mathcal{K}_{\infty}$ functions $\alpha_1,\alpha_2$ and class $\mathcal{K}$ function $\alpha_{3}$ such that, for all $x \in \mathcal{X}$,
	\begin{align}
		&\alpha_1(\|x\|) \leq V(x)  \leq \alpha_2(\|x\|), \label{CLF1} \\  
		&\inf _{u \in \Re^m} \left[ \frac{\partial V(x)}{\partial x}\left( f(x)+\phi {{(x)}^{T}}\theta+g(x)u \right) \right]  \leq-\alpha_3(\|x\|).  \label{CLF2}
	\end{align}
}

\vspace{-0.2cm}
To construct the CBFs, a set $\mathcal{C} \subset \mathcal{X}$ is defined as the zero superlevel set of a continuously differentiable function $h:\mathcal{X} \rightarrow \Re$ yielding:
\begin{align}
	\mathcal{C}&=\left\{x \in \mathcal{X} \subset \Re^n: h(x) \geq 0\right\}, \label{C1} \\  
	\partial \mathcal{C}&=\left\{x \in \mathcal{X} \subset \Re^n: h(x)=0\right\}, \label{C2} \\  
	\operatorname{Int}(\mathcal{C})&=\left\{x \in \mathcal{X} \subset \Re^n: h(x)>0\right\}.  \label{C3}
\end{align}
We refer to $\mathcal{C}$ as the safe set. 
The forward invariant property of $\mathcal{C}$ is defined as follows.  \\
\\
\text{\bf Definition 2 \citep{ames2017control}.}
\textit{Given a feedback controller for (\ref{model}), the set $\mathcal{C}$ is forward invariant for the resulting closed-loop system if $x(t) \in \mathcal{C}, \forall t >0$ for every $x(0) \in \mathcal{C}$.}\\
\\
It is obvious that the safety constraint is equivalent to the forward invariant property of the safe set $\mathcal{C}$, then the CBFs are defined as follows.  \\
\\
\text{\bf Definition 3 \citep{xu2015robustness}.}
\textit{The function $h$ in (\ref{C1})-(\ref{C3}) is a Zeroing Control Barrier Function (ZCBF) if there exists an extended class $\mathcal{K}$ function $\alpha_4$ such that, for all $x \in \mathcal{C}$,
\begin{align}
	\sup _{u \in \Re^m}\left[ \frac{\partial h(x)}{\partial x}\left( f(x)+\phi {{(x)}^{T}}\theta+g(x)u \right) \right] \geq-\alpha_4(h(x)). \label{ZCBF} 
\end{align}}

\vspace{-0.2cm}
From the existing results, most of the the control objectives and safety constraints can be transformed to CLFs and CBFs, respectively. 
For example, given a desired trajectory $x_d$, the tracking objectives can be achieved by establishing the tracking error $e=x-x_d$ and corresponding CLFs $V(e)$.
In order to construct a general adaptive safety control approach, this paper is not limited to specific control objectives and safety constraints.
	
For arbitrary state $x \in \mathcal{C}$, denote the sets of control inputs that satisfy (\ref{CLF2}) and (\ref{ZCBF}) as $\mathcal{U}_{\rm clf}(x)$ and $\mathcal{U}_{\rm cbf}(x)$, respectively. 
According to \citet{sontag1995characterizations,xu2015robustness,ames2017control}, the control input that belongs to  $\mathcal{U}_{\rm clf}(x)$ and $\mathcal{U}_{\rm cbf}(x)$ can achieve the control objectives and guarantee system safety, respectively.

\subsection{Problem Formulation}
In this paper, system (\ref{model}) is expected to safely achieve the control objectives, even if the control objectives and the safety constraints are tightly coupled and potentially conflicting. 
To complete the controller design and stability analysis, the following assumption is imposed.
\begin{assumption}
	System (\ref{model}) is initially relative safe, i.e. $h(x(0)) > \epsilon$ with a positive constant $\epsilon$.
\end{assumption}

\begin{remark}
	It is worth noting that in many existing results including \citet{xu2015robustness,ames2017control,taylor2020adaptive}, the initially safe assumption that $h(x(0)) \geq 0$ is imposed.  
	Considering the effects of system uncertainties in this paper, it has to be enhanced as $h(x(0)) > \epsilon$, for the design of safety triggered condition while avoiding the unexpected case that $h(x(0))=0$, $\dot{h}(x(0),u(0))<0$.
\end{remark}

It is worth noting that the system safety is always more important than other control objectives, i.e. the priority of (\ref{ZCBF}) is higher than that of (\ref{CLF2}).
If $\theta$ in (\ref{model}) is known, the control signal $u$ can be derived as in \citet{ames2014control,ames2017control,xu2015robustness}, by combining the CLFs (\ref{CLF2}) and CBFs (\ref{ZCBF}) with QP.
\begin{equation}
	\begin{aligned}
		u=&\underset{v \in \Re^{m+1}}{\mathop{\operatorname{argmin}}}\, \left(\frac{1}{2}{{v}^{T }}H(x)v+F{{(x)}^{T }}v\right)  \\ 
		s.t.&\frac{\partial V(x)}{\partial x}\left( f(x)+\phi {{(x)}^{T}}\theta+g(x)u \right)+{{\alpha }_{3}}\left( \left\| x \right\| \right)  \le \delta, \\ 
		&\frac{\partial h(x)}{\partial x}\left( f(x)+\phi {{(x)}^{T}}\theta+g(x)u \right)+{{\alpha }_{4}}(h(x))  \ge 0. \\ 	
	\end{aligned}
	\label{QP1}
\end{equation}
where $v=\left[u,\delta\right] \in \Re^{m+1}$. $\delta \in \Re$ is a relaxation term, $H(x) \in \Re^{(m+1) \times (m+1)}$ is positive definite, and $F(x) \in \Re^{m+1}$.
The CLFs and CBFs are conflicting when $\mathcal{U}_{\rm clf}(x) \cap \mathcal{U}_{\rm cbf}(x)=\emptyset$, then the relaxation term $\delta$ is calculated to be positive, to regrade the priority of CLF and guarantee the existence of solution to (\ref{QP1}).
However, it is not possible to calculate and implement the control input because $\theta$ in system (\ref{model}) is unknown. This is the main challenge addressed in this paper.

\section{Adaptive Safety Controller Design}
An adaptive law can be designed to generate the parameter estimate $\hat{\theta}(t)$, and the estimation error is defined as $\tilde{\theta}(t)=\theta-\hat{\theta}(t)$.
If the parameter update law can eliminate the effects of parameter uncertainties in finite time,
\begin{equation}
	\phi\left(x(t)\right)^T\tilde{\theta}(t)=0, \forall t \geq \tau_k,  \label{CE}
\end{equation}
where $\tau_{k}$ represents some finite time moment, then a certainty equivalence controller is derived by simply following (\ref{QP1}), i.e. 
\begin{equation}
	\begin{aligned}
		u=&\underset{v \in \Re^{m+1}}{\mathop{\operatorname{argmin}}}\, \left(\frac{1}{2}{{v}^{T }}H(x)v+F{{(x)}^{T }}v\right)  \\ 
		s.t.&\frac{\partial V(x)}{\partial x}\left( f(x)+\phi(x)^T \hat{\theta}(t)+g(x)u \right)+{{\alpha }_{3}}\left( \left\| x \right\| \right)  \le \delta, \\ 
		&\frac{\partial h(x)}{\partial x}\left( f(x)+\phi(x)^T \hat{\theta}(t)+g(x)u \right)+{{\alpha }_{4}}(h(x))  \ge 0. \\ 	
	\end{aligned}
	\label{QP2}
\end{equation}
With (\ref{CE}) and (\ref{QP2}), the safe set $\mathcal{C}$ is guaranteed to be forward invariant $\forall t \in \left[ \tau_{k}, \infty \right)$.
To guarantee the forward invariant property of $\mathcal{C}, \forall t \in \left[0,\tau_{k}\right)$, an adaptive update law with a safety triggered condition will be presented.
\begin{remark}
	Compared with most of the adaptive control results, the parameter estimate for the QP-based controllers should be more accurate, in the sense that $\phi\left(x(t)\right)^T \tilde{\theta}(t)=0$, $\forall t \geq \tau_k$.
	The main reason is that the sets $\mathcal{U}_{\rm clf}(x)$ and $\mathcal{U}_{\rm cbf}(x)$ may be miscalculated with the effects of estimation errors.
	An illustrating example is given in Fig. \ref{remark}, where (a) and (b) show the calculated $\mathcal{U}_{\rm clf}(x)$ and $\mathcal{U}_{\rm cbf}(x)$ with true $\theta$ and less accurate estimate $\hat{\theta}(t)$, respectively.
	The CLFs and CBFs that do not conflict with each other may be erroneously perceived as conflicting. 
	This will evidently affect the attainment of control objectives and safety constraints, further lead to conservative control performance (excessively guarantee safety constraints without achieving control objectives).
	\begin{figure}[!htbp]
		\vspace{-0.2cm}
		\centering
		\includegraphics[width=8cm]{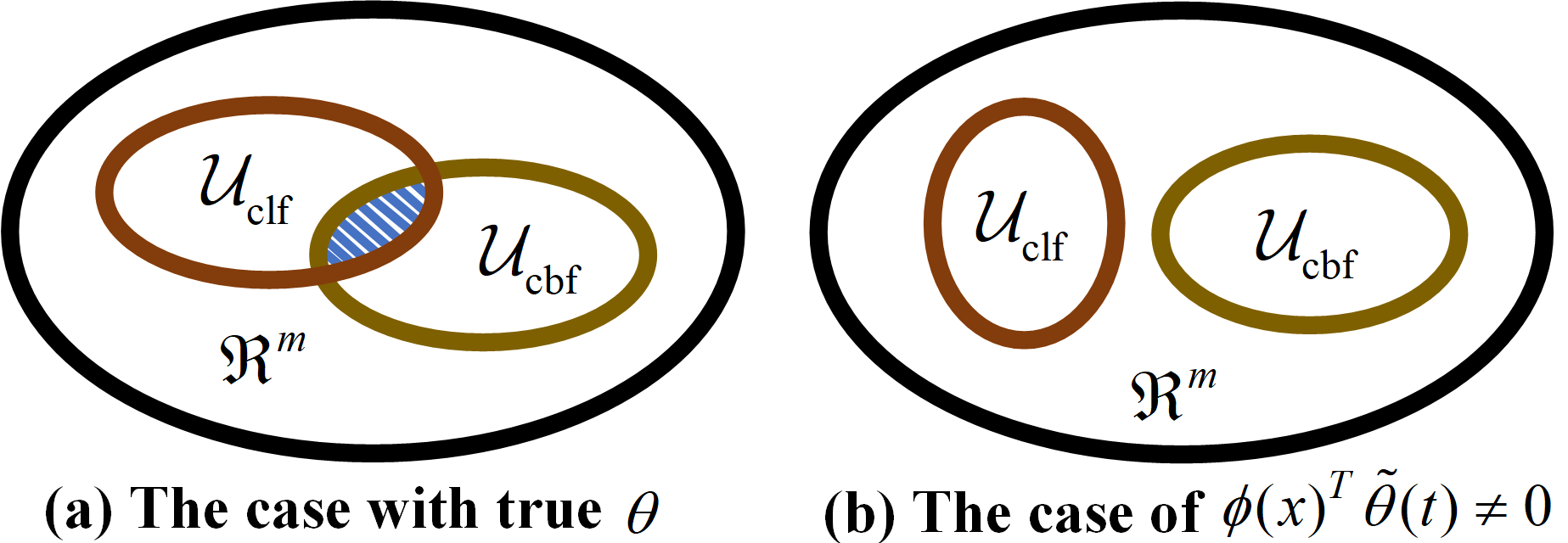}
		\vspace{-0.2cm}
		\caption{An illustrating example of the effect of less accurate estimate $\hat{\theta}(t)$ on $\mathcal{U}_{\rm clf}(x)$ and $\mathcal{U}_{\rm cbf}(x)$} 
		\label{remark}
		\vspace{-0.2cm}
	\end{figure}
\end{remark}
To achieve $\phi\left(x(t)\right)^T \tilde{\theta}(t)=0, \forall t \geq \tau_{k}$, a BaLSI-based parameter estimation method \citep{karafyllis2018adaptive,karafyllis2020adaptive} is presented in this paper for safety critical adaptive control. 
\subsection{Batch Least-Squares Identifier}
Similar to the adaptive control methods in \citet{ioannou1996robust},
the transformation is imposed to obtain the linear parametric models of (\ref{model}), which contain the information of the system inputs and outputs within a specific time interval.
Firstly, integrate system (\ref{model}) for arbitrary $t,\sigma \ge 0$,
\begin{equation}
	\begin{aligned}
		x(t)-x(\sigma)=
		&\int_{\sigma}^{t}{\bigg( f\big(x\left(s\right)\big)+g\big(x\left(s\right)\big)u\left(s\right) \bigg)ds }\\
		&+\left( \int_{\sigma}^{t}{\phi (x(s))^T}ds \right) \theta.
	\end{aligned}
	\label{para_1}
\end{equation}
Define for every $t,\sigma \ge 0$
\begin{align}
	p(t,\sigma)
	=&x(t)-x(\sigma) \nonumber \\ 
	&- \int_{\sigma}^{t}{\bigg( f\big(x\left(s\right)\big)+g\big(x\left(s\right)\big)u\left(s\right) \bigg)ds }, 
	\label{para_2}\\
	q(t,\sigma)
	=& \int_{\sigma}^{t}{\phi (x(s))^T}ds .
	\label{para_3} 
\end{align}
Equation (\ref{para_1}) can thus be rewritten as
\begin{equation}
	p(t,\sigma)=q(t,\sigma) \theta.
	\label{para_4}
\end{equation}
Then, the following equation can be obtained by integrating $q(t,\sigma)^Tp(t,\sigma)=q(t,\sigma)^Tq(t,\sigma) \theta$ with $t$ and $\sigma$.
\begin{equation}
	Z(\tau_{i})=G(\tau_{i}) \theta,
	\label{para_5}
\end{equation}
where $\tau_{i}$, $\forall i \in Z^+$ is a group of time moments, and
\begin{align}
	G\left(\tau_{i}\right)=\int_{0}^{\tau_{i}} \int_{0}^{\tau_{i}} q(t, \sigma)^T q(t, \sigma) d \sigma d t,
	\label{para_6} \\
	Z\left(\tau_{i}\right)=\int_{0}^{\tau_{i}} \int_{0}^{\tau_{i}} q(t, \sigma)^T p(t, \sigma) d \sigma d t.
	\label{para_7} 
\end{align}
Clearly, if $G(\tau_{i})$ is invertible, the unknown vector $\theta$ can be calculated as follows,
\begin{equation}
	\theta=G(\tau_{i})^{-1}Z(\tau_{i}).
	\label{para_8}
\end{equation}
However, since $G(\tau_{i}) \in \Re^{p \times p}$ is only symmetric and positive semidefinite, it is not guaranteed to be invertible.
Therefore  (\ref{para_8}) cannot be utilized to estimate the unknown vector $\theta$.
To solve this problem, the following quadratic optimization based parameter update law is proposed:
\begin{equation}
	\begin{aligned}  
		&\hat{\theta}\left(\tau_{i}\right)=\underset{\vartheta \in \Re^p}{\mathop{\operatorname{argmin}}}\, \left\{  \left \Vert \vartheta-\hat{\theta}\left(\tau_{i-1}\right) \right \Vert ^2  \right\}   \\
		&s.t. \quad Z\left(\tau_{i}\right)=G\left(\tau_{i}\right) \vartheta.
	\end{aligned} 
	\label{para_9}
\end{equation}
The estimate $\hat{\theta}(t)$ is updated at $\tau_{i}$ by utilizing the stored data during the time interval $\left[0,\tau_i\right), \forall i \in Z^+$, and it is kept unchanged between any two successive update moments.
\begin{equation}
	\hat{\theta}(t)=\hat{\theta}(\tau_i), \forall t \in [\tau_i,\tau_{i+1}).
\end{equation}
As discussed in \citet{karafyllis2018adaptive}, (\ref{para_9}) implies that $\hat{\theta}(\tau_{i})$ is the projection of $\hat{\theta}(\tau_{i-1})$ on the hyperplane defined by the linear equality constraints  $Z\left(\tau_{i}\right)=G\left(\tau_{i}\right) \vartheta$.
The parameter update law utilizes double integrals in (\ref{para_6}) and (\ref{para_7}) rather than $\tilde{G}\left(\tau_{i}\right)=\int_{0}^{\tau_{i}} q(t, 0)^T q(t, 0) d t$ and $\tilde{Z}\left(\tau_{i}\right)=\int_{0}^{\tau_{i}} q(t, 0)^T p(t, 0) d t$, since it gives equal weight to all measurements.
Note that $G\left(\tau_{i}\right)$ and $Z\left(\tau_{i}\right)$ can be obtained via a group of ODEs, which are omitted in this paper. Interested readers may refer to \citet{karafyllis2020adaptive}.

\subsection{Safety Triggered Condition}
Note that the parameter update law (\ref{para_9}) should not be activated at each sampling time moment, since it may lead to the undesired discontinuous parameter estimation. 
To complete the design and analysis, the following definition and assumption are presented

\text{\bf Definition 4 \citep{Heymann2005}.}
The parameter update law (\ref{para_9}) is Zeno if infinite number of updates occur within a finite time interval.

\begin{assumption}
	For the extended class $\mathcal{K}$ function $\alpha_{4}$ in (\ref{ZCBF}), there exists two positive constants  $\overline{\chi} \in \Re^+$, $\sigma \in \Re^+$ such that $\frac{x}{\alpha_{4}(x)} \ge \sigma, \forall x \in \left(0,\overline{\chi}\right]$.
\end{assumption}

\begin{remark}
	Assumption 2 is imposed to exclude the Zeno behavior of the proposed adaptive update law.
	Detailed analysis will be given in Section 4.
	Note that Assumption 2 is not excessively strict, since many extended class $\mathcal{K}$ functions can satisfy it.
	Considering a common case $\alpha_{4}(x)=\kappa x^\lambda$, where $\kappa \in \Re^+$ and  $\lambda$ is a positive odd number, we have $\frac{x}{\alpha_{4}(x)}=\frac{1}{\kappa}x^{1-\lambda}$.
	It can be easily seen that $\frac{x}{\alpha_{4}(x)} \ge \sigma, \forall x \in \left(0,\overline{\chi}\right]$ with $\overline{\chi}=\left( \sigma \kappa \right)^ {\frac{1}{1-\lambda}} $.
\end{remark}

Due to the fact that system safety is more important than control objectives, the proposed triggered condition that determines the updating moments $\tau_i$, should be related to the degree of system safety.
Consider the situation that $\exists t \in \left[\tau_{i-1},\tau_{i}\right)$, $\phi\left(x(t)\right)^T \tilde{\theta}(t)$  $\neq 0$, indicating that control performance is affected by inaccurate parameter estimate. 
To guarantee the forward invariant property of the safe set $\mathcal{C}$, the estimate $\hat{\theta}(t)$ should be updated before the system state $x$ reaches the boundary of the safe set, $\partial\mathcal{C}$.
According to (\ref{C1})-(\ref{C3}), CBF $h$, which can directly reflect the degree of system safety, can be adopted as safety-triggered function.
Then the safety-triggered moment can be determined if the following condition holds,
\begin{equation}
	h\left(x(\tau_i)\right) = \chi_{i},
	\label{trigger_1}
\end{equation}
where $\chi_{i}>0$, $i \in Z^+$ denotes the alarm value of each triggering moment.
To avoid continuous triggering of (\ref{para_9}), $\chi_{i}$ is chosen as
\begin{equation}
	\begin{aligned}  
		\chi_{1}&=\min \left\{\overline{\chi},\gamma_0 h\left(x(0)\right) \right\},  \\
		\chi_{i}&=\gamma_1 \chi_{i-1}, \forall i \ge 2,
	\end{aligned}
	\label{trigger_2}
\end{equation}
where $0<\gamma_0<1$ and $0<\gamma_1<1$ are design parameters of the safety-triggered condition, which are used to decrease the value of $\chi_i$ at each triggering moment.
The terms $\overline{\chi}$ and $\gamma_0 h\left(x(0)\right)$ are involved in the $\min\left\{ \bullet, \bullet \right\}$ function of (\ref{trigger_2}) to exclude the Zeno behavior of the proposed adaptive update law.
Detailed analysis will be given in Section 4.

Note that condition (\ref{trigger_1}) may not be satisfied due to the effects of inaccurate estimate, i.e. $\phi\left(x(t)\right)^T \tilde{\theta}(t) \neq 0$, while the affected system is kept relatively safe.
Although the safety constraints are not violated, the control objectives may not be achieved with inaccurate estimate, especially for the case that CLFs and CBFs do not conflict with each other whereas they are erroneously perceived as conflicting, as explained in Fig. \ref{remark}.
Hence, the effects of unknown parameters are expected to be totally eliminated, i.e. $\phi\left(x(t)\right)^T\tilde{\theta}(t)=0, \forall t \geq \tau_k$.
We define a constant $\Delta \tau_{\max} \in \Re^+$, in which the parameter estimate should be updated at least once.
Combining  $\Delta \tau_{\max}$ and (\ref{trigger_1}), the triggered moment $\tau_{i}$ is formally described as
\begin{equation}
	\tau_{i}=\min \bigg\{ \big\{ t>\tau_{i-1}:h\left(x(t)\right) = \chi_{i} \big\} ,  \tau_{i-1}+\Delta \tau_{\max} \bigg\}.
	\label{trigger_3}
\end{equation}
To avoid the confused representation of $\tau_{i-1}$ for $i=1$, we state that $\tau_0=0$.
Now, the proposed adaptive control scheme is completed, with the QP-based controller (\ref{QP2}), the BaLSI-based adaptive update law (\ref{para_9}) and the safety-triggered condition (\ref{trigger_3}).

\section{Stability Analysis}
Before analyzing the stability of the closed-loop system, some properties are firstly established in the following lemma.
\begin{lemma}
	With the proposed parameter update law (\ref{para_9}), (\ref{trigger_3}), the updated parameter estimate $\hat{\theta}(\tau_{i})$ satisfies $\phi\left(x(t)\right)^T \tilde{\theta}(\tau_{i})=0$, $\forall t \in \left[0,\tau_{i}\right)$, $\forall i \in Z^+$.
\end{lemma}
\text{\bf Proof.}
	According to the proposed adaptive update law (\ref{para_9}), it is obvious that,
	\begin{equation}
		Z\left(\tau_{i}\right)=G\left(\tau_{i}\right) \hat{\theta}(\tau_{i}).
		\label{lemma1_1}
	\end{equation}
	From (\ref{para_5}) and (\ref{lemma1_1}), we have
	\begin{equation}
		G\left(\tau_{i}\right) \tilde{\theta}(\tau_{i}) = 0,
		\label{lemma1_2}
	\end{equation}
	where $\tilde{\theta}(\tau_{i})$ is
	\begin{equation}
		\tilde{\theta}(\tau_{i})=\theta - \hat{\theta}(\tau_{i}).
		\label{lemma1_3}
	\end{equation}
	From the definition of $G\left(\tau_{i}\right)$ in (\ref{para_6}), and multiplying (\ref{lemma1_2}) from the left with $\tilde{\theta}(\tau_{i})^T$, we have
	\begin{equation}
		\int_{0}^{\tau_{i}} \int_{0}^{\tau_{i}} \left\Vert q(t, \sigma) \tilde{\theta}(\tau_{i}) \right\Vert ^2 d \sigma d t = 0.
		\label{lemma1_4}
	\end{equation}
	From (\ref{para_3}) and (\ref{lemma1_4}), it is obvious that $\phi\left(x(t)\right)^T \tilde{\theta}(\tau_{i})=0,\forall t \in \left[0,\tau_{i}\right)$.
	\hfill$\blacksquare$

\begin{remark}
	From Lemma 1 and the fact that the regressor $\phi(x)$ represents excitation information, the proposed adaptive update law (\ref{para_9}) utilizes all the excitation information during $[0,\tau_{i})$ to update the parameter estimate $\hat{\theta}(\tau_{i})$.
	The situation $\exists t \in \left[\tau_{i},\tau_{i+1}\right)$, $\phi(x(t))^T\tilde{\theta}(\tau_{i}) \neq 0$ indicates that there exists some new excitation information during the interval.
	In other words, it cannot be expressed as a linear combination of the regressors $\phi(x(t))$ during $[0,\tau_i)$.
	Then the parameter estimate will be updated to guarantee $\phi\left(x(t)\right)^T \tilde{\theta}(\tau_{i+1})=0$, $\forall t \in \left[0,\tau_{i+1}\right)$.
\end{remark}

Note that the parameter estimate may not be updated at the triggering moment, hence the times of estimate updating is not equal to the times of triggering.
The main results of this paper are formally stated in the following theorems. 
 
\begin{theorem}
	Consider the system modeled by (\ref{model}) with initial condition under Assumption 1.
	With the proposed control scheme  (\ref{QP2}), (\ref{para_9}), (\ref{trigger_3}), the following holds.\\
		(i)
		The parameter estimate will be updated by (\ref{para_9}) no more than $p$ times, where $p$ is dimension of $\theta$.
		Denote $\tau_k$ as the last moment at which the estimate is updated, the effects of unknown parameter are eliminated after $\tau_k$, in the sense that $\phi(x(t))^T \tilde{\theta}(\tau_k)=0$ for all $t \geq \tau_k$.\\
		(ii)
		Function $h$ is guaranteed to be non-negative along solutions of the closed-loop system, which indicates that safety constraints are guaranteed.
\end{theorem}

\text{\bf Proof.}
	(i)
	Denoting the range space and null space of matrix $G(\tau_i)$ as $\mathcal{R}[G(\tau_i)]$ and $\mathcal{N}[G(\tau_i)]$ respectively, there is $\dim \mathcal{R}[G(\tau_i)]+\dim \mathcal{N}[G(\tau_i)]=p$.
	When condition (\ref{trigger_3}) is triggered and $\tilde{\theta}(\tau_{i}) \notin \mathcal{N}[G(\tau_{i+1})]$, the parameter estimate should be updated to guarantee $\tilde{\theta}(\tau_{i+1}) \in \mathcal{N}[G(\tau_{i+1})]$.
	From Remark 4, there exists new excitation information during the interval $[\tau_{i},\tau_{i+1})$, i.e. $\dim \mathcal{R}[G(\tau_{i+1})] \geq \dim \mathcal{R}[G(\tau_{i})]+1$.
	It can be further concluded that $\dim \mathcal{N}[G(\tau_{i+1})] \le \dim \mathcal{N}[G(\tau_{i})]-1$.
	According to the monotone bounded convergence theorem, the dimension of $\mathcal{N}[G(\tau_{i})]$ will eventually decrease to a non-negative integer, and the parameter estimate will be updated no more than $p$ times.
	Furthermore, condition (\ref{CE}) is obviously satisfied, otherwise $\hat{\theta}$ has to be updated again according to the new excitation information, which violates the statement that $\tau_{k}$ is the last updating moment.
	Hence the effects due to unknown system parameters will be totally eliminated $\forall t \ge \tau_{k}$.
	
	(ii)
	For a time interval between two successive triggering moments, if $\exists t \in \left[\tau_{i},\tau_{i+1}\right)$, $\phi\left(x(t)\right)^T \tilde{\theta}(\tau_{i}) \neq 0$ and the inaccurate estimates pull the state $x$ towards $\partial \mathcal{C}$, condition (\ref{trigger_3}) will be triggered before $x$ escapes from the safe set $\mathcal{C}$, indicating that $h(x(t))>0, \forall t \in \left[\tau_{i},\tau_{i+1}\right)$.
	If the estimate satisfies $\phi\left(x(t)\right)^T\tilde{\theta}(\tau_{i})=0, \forall t \in \left[\tau_{i},\tau_{i+1}\right)$, the performance of the adaptive controller (\ref{QP2}) will be totally the same with the case that $\theta$ is known, and $h(x(t))>0, \forall t \in \left[\tau_{i},\tau_{i+1}\right)$ can be concluded by utilizing similar proof in \citet{xu2015robustness,ames2017control}.
	That is to say, the inequality $h(x(t)) \geq 0$ for $\exists t \in \left[\tau_{i},\tau_{i+1}\right), \phi\left(x(t)\right)^T\tilde{\theta}(\tau_{i}) \neq 0$ and $\phi\left(x(t)\right)^T\tilde{\theta}(\tau_{i})=0, \forall t \in \left[\tau_{i},\tau_{i+1}\right)$ are guaranteed by the triggered condition (\ref{trigger_3}) and CBFs, respectively.
	Thus the proposed control scheme guarantees the safety constraint along solutions of the closed-loop system.
	$\hfill\blacksquare$
	
	\begin{remark}
		From the definition of $G(\tau_i)$ in (\ref{para_6}), it is a symmetric matrix.
		According to matrix theory, the spaces $\mathcal{R}[G(\tau_i)]$ and $\mathcal{N}[G(\tau_i)]$ are orthogonal complement of each other, i.e. $\Re^p=\mathcal{R}[G(\tau_i)] \bigoplus \mathcal{N}[G(\tau_i)]$.
		From Lemma 1, $\mathcal{R}[G(\tau_i)]$ involves all the previously appeared excitation information, and the parameter estimation error $\tilde{\theta}(\tau_i)$ belongs to $\mathcal{N}[G(\tau_i)]$.
		Thus $\mathcal{R}[G(\tau_i)]$ and $\mathcal{N}[G(\tau_i)]$ can be seen as excited subspace and unexcited subspace of the estimation error space $\Re^p$, respectively.
		The orthogonal projections of the initial estimation error $\tilde{\theta}(0)$ on $\mathcal{R}[G(\tau_i)]$ and $\mathcal{N}[G(\tau_i)]$ are defined as excited and unexcited components.
		The excited component of the estimation error is eliminated at the updating moments, while the unexcited component is kept unchanged.
	\end{remark}
	
	From Remark 4 and Remark 5, $\hat{\theta}(t)$ may be kept unchanged when condition (\ref{trigger_3}) is triggered.
	The triggers are divided into the following three cases, while only one of them will cause the change of parameter estimate. \\
	{\bf Case 1.}
		There is no new excitation information in the time interval, in the sense that $\phi(x(t))^T\tilde{\theta}(\tau_{i-1})=0$, $\forall t \in \left[\tau_{i-1},\tau_i\right)$.
		In this case, condition (\ref{trigger_3}) is triggered by $\tau_{i}=\tau_{i-1}+\Delta \tau_{\max}$, thus $\hat\theta(t)$  will not be updated for $t=\tau_i$.\\
	{\bf Case 2.}
		Note that (\ref{QP2}) allows $h(x)$ to be slowly decreased, $\forall x \in \mathcal{C}$.
		Hence (\ref{trigger_3}) may be triggered at the moment $\tau_{i} \leq \tau_{i-1}+\Delta \tau_{\max}$ with no new excitation information.
		In this case, the triggering is caused by the conflict between control objectives and safety constraints, hence $\hat\theta(t)$ will not be updated for $t=\tau_i$.\\
	{\bf Case 3.}
		There exists new excitation information in the time interval, in the sense that $\exists t \in \left[\tau_{i-1},\tau_i\right)$, $\phi(x(t))^T\tilde{\theta}(\tau_{i-1}) \neq 0$.
		The control performance is affected by inaccurate parameter estimate, which may cause the decreasing of $h(x)$ and trigger condition (\ref{trigger_3}).
		The excitation leads to $G\left(\tau_{i}\right) \tilde{\theta}(\tau_{i-1}) \neq 0$, and $\hat\theta(t)$ is updated to satisfy (\ref{lemma1_2}).

\begin{figure}[!htbp]
	\vspace{-0.2cm}
	\centering
	\includegraphics[width=7cm]{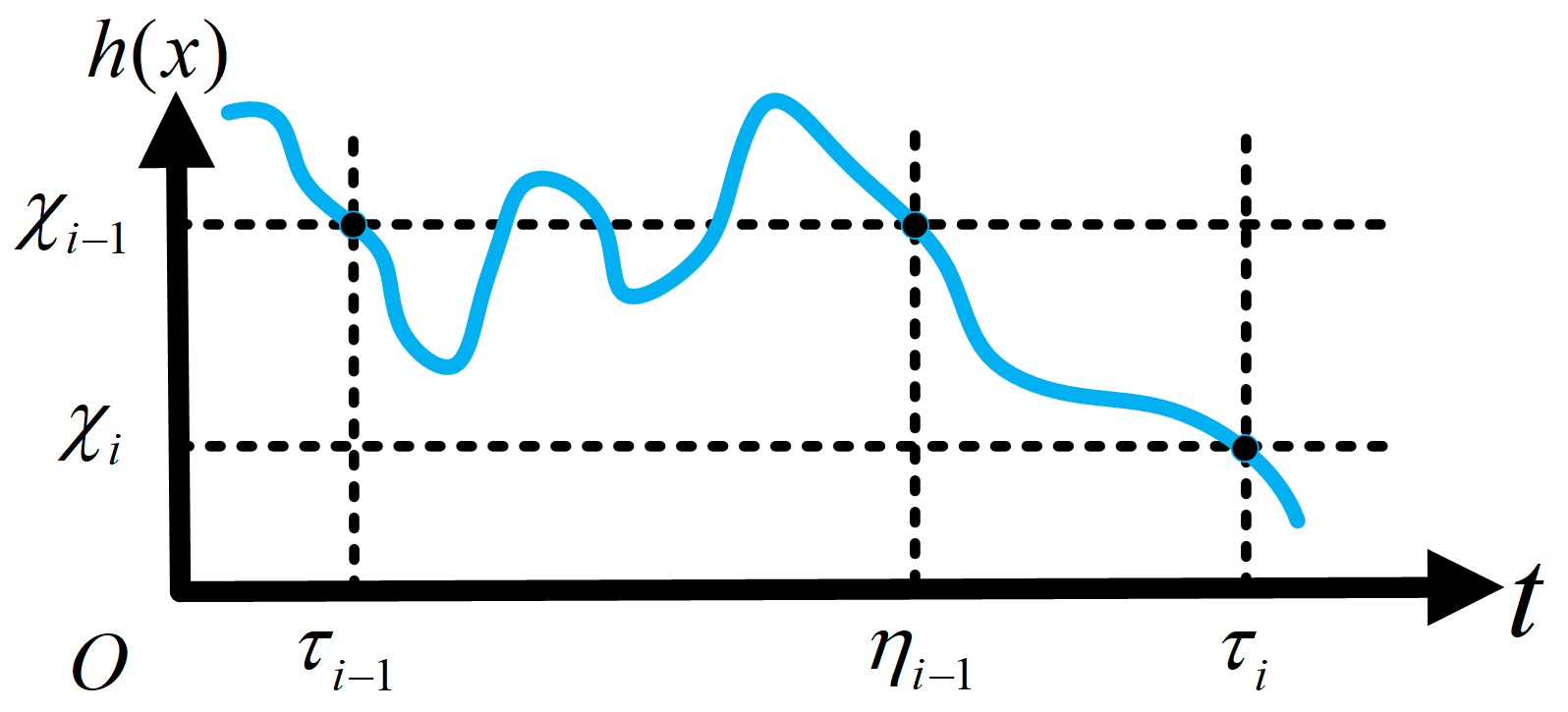}
	\vspace{-0.4cm}
	\caption{An illustrating example of time moments} 
	\label{time}
	\vspace{-0.2cm}
\end{figure}

\begin{theorem}
	Consider system (\ref{model}) with a CBF under Assumptions 1-2.
	With the proposed control scheme  (\ref{QP2}), (\ref{para_9}), (\ref{trigger_3}), Zeno behavior of the parameter update law (\ref{para_9}) can be excluded.
\end{theorem}
\text{\bf Proof.}
Note that $\tau_{i}-\tau_{i-1}=\Delta \tau_{\max}$ in Case 1 is a positive constant and the times that (\ref{trigger_3}) triggered by Case 3 is less than $p$, which indicates that Cases 1 and 3 will not lead to Zeno behavior.
To analyze the infimum  of $\tau_{i}-\tau_{i-1}$ in Case 2, denote $\eta_{i-1}$ as the moment such that $\max \left\{t \in \left[\tau_{i-1}, \tau_{i}\right) : h(x(t))=\chi_{i-1} \right\}$.
Fig. \ref{time} is given to explain the relationship between $\tau_{i-1}$, $\eta_{i-1}$ and $\tau_{i}$.

Integrate the inequality $\dot{h} \ge -\alpha_{4}(h)$ with the time interval $\left[\eta_{i-1},\tau_{i}\right]$, we have
\begin{equation}
	h(x(\tau_{i}))-h(x(\eta_{i-1})) \ge - \int_{\eta_{i-1}}^{\tau_{i}} \alpha_{4}\big(h(x(t))\big) dt. 
	\label{proof4_1}
\end{equation}
Substituting (\ref{trigger_1}) and (\ref{trigger_2}) into (\ref{proof4_1}), we have
\begin{equation}
	\begin{aligned}  
		(\gamma_1-1)\chi_{i-1}
		&=h(x(\tau_{i}))-h(x(\eta_{i-1})) \\
		&\ge - \int_{\eta_{i-1}}^{\tau_{i}} \alpha_{4}\big(h(x(t))\big) dt  \\
		&\ge - (\tau_{i}-\eta_{i-1}) \alpha_{4}\big(\chi_{i-1}\big).
	\end{aligned}
	\label{proof4_11}
\end{equation}
Then according to Assumption 2 and $\chi_{i-1}>0$, we get
\begin{equation}
	\begin{aligned}
		\tau_{i}-\tau_{i-1} 
		&\ge \tau_{i}-\eta_{i-1} \\
		&\ge (1-\gamma_1) \frac{\chi_{i-1}}{\alpha_{4}\big(\chi_{i-1}\big)} \\
		&\ge (1-\gamma_1) \sigma,
	\end{aligned}
	\label{proof4_2}
\end{equation}
where $(1-\gamma_1) \sigma$ is a positive constant, whose value can be adjusted by properly choosing the extended class $\mathcal{K}$ function $\alpha_{4}$ in (\ref{ZCBF}) and the relaxing coefficient $\gamma_1$ in (\ref{trigger_2}).
Thus Zeno behavior can be excluded for the adaptive update law (\ref{para_9}).
$\hfill\blacksquare$

\begin{remark}
	Compared with \citet{karafyllis2018adaptive,karafyllis2020adaptive}, the conflicting control objectives and safety constraints are considered in this paper.
	The proposed triggered condition is different, due to the fact that the CBFs  \citep{xu2015robustness,ames2017control} defined in (\ref{ZCBF}) are allowed to decrease to mediate the potential conflict.
	Condition (\ref{trigger_3}) may be triggered by both inaccurate estimate and conflict between the control objectives and safety constraints, thus the stability analysis on safety is more challenging.
	Assumption 2 and the triggered condition (\ref{trigger_2}) are presented to exclude the Zeno behavior.
\end{remark}

\begin{remark}
	The Lyapunov-based adaptive control laws proposed for (\ref{QP2}) in \citet{taylor2020adaptive} are given below,
	\begin{align}
		\dot {\hat{\theta}}_V &= \Gamma_1 \left(\frac{\partial V_a (x,\hat{\theta}_V)}{\partial x} \phi(x)^T \right)^T,
		\label{Lyapunov_1} \\  
		\dot {\hat{\theta}}_h &= -\Gamma_2 \left(\frac{\partial h_a (x,\hat{\theta}_h)}{\partial x} \phi(x)^T \right)^T, \label{Lyapunov_2} 
	\end{align}
	where $V_a(x,\hat{\theta}_V)=V(x)+\tilde{\theta}_V^T \Gamma_1^{-1} \tilde{\theta}_V$ is the adaptive Control Lyapunov Function (aCLF), $h_a(x,\hat{\theta}_h)=h(x)-\tilde{\theta}_h^T \Gamma_2^{-1} \tilde{\theta}_h$ is the adaptive Control Barrier Function (aCBF).
	$\tilde{\theta}_V(t)=\theta-\hat{\theta}_V(t)$ and $\tilde{\theta}_h(t)=\theta-\hat{\theta}_h(t)$ are parameter estimation errors for aCLFs and aCBFs, respectively.
	$\Gamma_1$ and $\Gamma_2$ are positive definite matrices.
	According to related research works on aCLFs \citep{krstic1995control}, the adaptive law in (\ref{Lyapunov_1}) can estimate the unknown parameters for single aCLF.
	However, the direct utilization of (\ref{Lyapunov_1}) and (\ref{Lyapunov_2}) to update the estimates for (\ref{QP2}) without any modification, will lead to the following disadvantages.
	\begin{enumerate}
		\item 
		The relaxation term $\delta$ in (\ref{QP2}), which is added to mediate the potential conflict between control objectives and safety constraints, will influence the asymptotic stability of $V_a(x,\hat{\theta}_V)$.
		In fact, the boundedness of parameter estimate cannot be derived from the Lyapunov function which yields only $\dot V_a(x,\hat{\theta}_V) \le -\alpha(\|x\|)+\delta$, with $\alpha$ being a class $\mathcal{K}$ function.
		The estimation error may grow unboundedly, which will affect the calculated control signal in (\ref{QP2}) and lead to an undesired control performance.
		\item 
		The adaptive law (\ref{Lyapunov_2}) for aCBFs can only ensure $h(x) \ge \tilde{\theta}_h^T \Gamma_2^{-1} \tilde{\theta}_h \ge 0$ rather than the boundedness of $\tilde{\theta}_h(t)$.
		When the estimation error increases, $h(x) \ge \tilde{\theta}_h^T \Gamma_2^{-1} \tilde{\theta}_h$ will keep the state $x$ within a quite safe subset of $\mathcal{C}$, which is denoted as $\Omega_{\tilde{\theta}_h}=\left\{x \in \mathcal{X} :h(x) \ge \tilde{\theta}_h^T \Gamma_2^{-1} \tilde{\theta}_h\right\}$. 
		In other words, the control objectives for all $x \in \mathcal{C}-\Omega_{\tilde{\theta}_h}$ cannot be achieved.
		The main aim of the QP-based controller (\ref{QP2}), mediating the potential conflict between control objectives and safety constraints, is broken by inaccurate estimates.
		\item 
		The number of parameter estimates is greater than the number of unknown parameters, thus the adaptive scheme in (\ref{Lyapunov_1}) and (\ref{Lyapunov_2}) is overparametrizated.
	\end{enumerate}
	Note that the above three disadvantages will not occur in the proposed adaptive scheme (\ref{para_9}), (\ref{trigger_3}).
\end{remark}

\begin{remark}	
	As discussed in \citet{lopez2020robust}, the unbounded estimates of the Lyapunov-based adaptive laws for (\ref{QP2}) may be avoided by employing the Lipschitz continuous parameter projection operators, which can be found in Appendix E of \citet{krstic1995nonlinear}.
	It is assumed that there exists a pre-known closed convex set $\Theta \subset \Re^p$ satisfying $\theta \in \Theta$.
	Then the adaptive laws
	can guarantee both $\dot V_a(x,\hat{\theta}_V) \le -\alpha(\|x\|)+\delta$ and $\hat{\theta}_V(t), \hat{\theta}_h(t) \in \Theta, \forall t \ge 0$, which result in the boundedness of all the closed-loop signals.
	Several novel data-driven adaptive control schemes, including the set membership identification based adaptive control \citep{lopez2020robust,lopez2023unmatched}, the concurrent learning based adaptive control \citep{isaly2021adaptive,cohen2022high} and the tightening parameter bounds based adaptive control \citep{wang2024adaptive}, have been proposed to reduce the conservation of the control performance, in the sense that range of the forward invariant set $\Omega_{\tilde{\theta}_h}=\left\{x \in \mathcal{X} :h(x) \ge \tilde{\theta}_h^T \Gamma_2^{-1} \tilde{\theta}_h\right\}$ will increase when the bounds of $\| \tilde{\theta}_h\|$ decrease.
	However, the controller (\ref{QP2}) with the projection operators aided adaptive laws or the data-driven adaptive laws 
	still works worse than (\ref{QP1}), with $\theta$ being known, due to the fact that (\ref{CE}) cannot be achieved.
	Moreover, all the mentioned adaptive control schemes robustly enforce the CBF condition using certain pre-known closed convex set $\Theta \subset \Re^p$, which leads their control performances to be highly conservative.
\end{remark}
\begin{remark}
	Compared with the classical estimation-based adaptive control methods in 
	\citet{ioannou1996robust}, the requirements on persistent excitation can be removed by the triggered BaLSI-based adaptive methods \citep{karafyllis2018adaptive,karafyllis2020adaptive}.
	We get $\tilde{\theta}(\tau_{i}) \in \mathcal{N}[G(\tau_{i})]$ in the proof of Lemma 1, which indicates that the previous excitation can be sufficiently utilized by the proposed adaptive update law (\ref{para_9}).
	According to $\phi\left(x(t)\right)^T \tilde{\theta}(\tau_{k})=0,\forall t \in [0,\tau_k)$, the unknown system parameters can be accurately estimated under sufficient excitation rather than persistent excitation.
	Moreover, the closed-loop system can still safely achieve control objectives even if the excitation is not sufficient, since the unexcited component of the estimation error will never affect the system (\ref{model}), i.e. $\tilde{\theta}(\tau_{k}) \neq 0$ but $\phi\left(x(t)\right)^T\tilde{\theta}(\tau_{k})=0, \forall t \ge 0$.
	Hence, with the proposed adaptive update law (\ref{para_9}), no assumption needs to be imposed on the initial parameter estimation value and excitation information.
\end{remark}

\section{Simulation Results}
To demonstrate the effectiveness of the proposed control scheme, the Adaptive Cruise Control (ACC) problem in \citet{taylor2020adaptive,wieland2007constructive} is considered.
The vehicle is modeled by,
\begin{align}
	\dot x_f &=v_f,   \label{follower_k} \\  
	\dot v_f &=\frac{1}{M} \tau_f+\phi(v_f)^T\theta,   \label{follower_d} 
\end{align}
where $x_f$, $v_f$ and $\tau_f$ denote the position, velocity and control torque of the following vehicle, respectively.
$M$ denotes the mass of vehicle, which is supposed to be known.
$\phi : \Re \rightarrow \Re^{3}$ is a known function and $\theta \in \Re^3$ is an unknown vector.
The uncertain term $\phi(v_f)^T\theta$ represents the vehicle's unmodeled dynamics, road frictional resistance and aerodynamic drag.

The control objective is to drive the velocity of the following vehicle to a desired velocity, $v_d=20 (\text{m/s})$, while ensuring that its distance to the forward vehicle satisfies a safety constraint:
\begin{equation}
	D \triangleq x_l-x_f \ge k_d v_f,
	\label{condition1}
\end{equation} 
where $x_l$ is the position of the forward vehicle, $k_d$ is the desired time headway \citep{wieland2007constructive}.
The velocity of the forward vehicle is chosen as,
\begin{equation}
	v_l(t)= 
	\begin{cases}
		18,    & {\rm if} \text{ } 0     \leq t < 5   \\ 
		18-4(t-5),             & {\rm if} \text{ } 5   \leq t < 7    \\
		10, & {\rm if} \text{ } 7    \leq t < 30 \\ 
		10+2(t-30),            & {\rm if} \text{ } 30 \leq t < 35    \\
		20,            & {\rm if} \text{ } 35 \leq t < 60    \\
	\end{cases},
	\label{leader}
\end{equation}
where $v_l$ during the time intervals $[5\text{s},7\text{s})$ and $[30\text{s},35\text{s})$ are chosen to simulate abrupt braking and aggressive acceleration of the forward vehicle.
Note that the desired velocity of the following vehicle is quicker than the forward vehicle, i.e. $v_d(t) \ge v_l(t)$, thus the control objective and safety constraint are conflicting.

The CLF and CBF are respectively chosen as
\begin{align}
	V &=\frac{1}{2}\left(v_f-v_d\right)^2,  \label{CLF_s} \\  
	h &=x_l-x_f - k_d v_f.   \label{CBF_s} 
\end{align}
The QP-based controller is given as,
\begin{equation}
	\begin{aligned}
		\tau_f=&\underset{v  \in  \Re^2}{\mathop{\operatorname{argmin}}}\, 
		\left(\frac{1}{2}{v^T}H(x)v+F(x)^T v\right)  \\ 
		s.t.
		&\frac{v_f-v_d}{M} \tau_f -\delta \le -\left(v_f-v_d\right)\phi^T \hat \theta(t) -k_1 V(x),\\ 
		&\frac{k_d}{M} \tau_f \le \left(v_l-v_f\right)-k_d \phi^T \hat \theta(t) +k_2 h(x).
	\end{aligned}
	\label{QP_ACC}
\end{equation}
where $v=\left[\tau_f,\delta\right] \in \Re^{2}$.

To implement the controller (\ref{QP_ACC}), four cases with different adaptive laws are considered sequentially.\\
\textbf{Case I.} The Lyapunov-based adaptive laws in \citet{taylor2020adaptive} are adopted, i.e. 
\begin{align}
	\dot {\hat{\theta}}_V &= c_1 \left(\frac{\partial V}{\partial v_f} \phi(v_f)^T \right)^T
	= c_1 \left(v_f-v_d\right) \phi(v_f),                \label{Lyap_1} \\  
	\dot {\hat{\theta}}_h &= -c_2 \left(\frac{\partial h}{\partial v_f} \phi(v_f)^T \right)^T
	= c_2 k_d\phi(v_f), \label{Lyap_2} 
\end{align}
where $c_1$ and $c_2$ are positive constants.\\
\textbf{Case II.} The Lyapunov-based adaptive laws with the Lipschitz continuous parameter projection operators in Remark 8  are utilized,
\begin{align}
	\dot {\hat{\theta}}_V &
	= {\rm Proj} \big( c_1 \left(v_f-v_d\right) \phi(v_f) \big),       \label{Lyap_proj_1} \\  
	\dot {\hat{\theta}}_h &
	= {\rm Proj} \big(c_2 k_d\phi(v_f) \big),  \label{Lyap_proj_2} 
\end{align}
where the pre-known closed convex set of $\theta$ is chosen as $\Theta_1=\left[0.8\theta,1.2\theta\right] \subset \Re^3$.\\
\textbf{Case III.} The adaptive laws (\ref{Lyap_proj_1}) and (\ref{Lyap_proj_2}) are adopted, the pre-known close convex set is chosen as $\Theta_2=\left[0.2\theta,1.8\theta\right] \subset \Re^3$.\\
\textbf{Case IV.} The proposed BaLSI-based adaptive law (\ref{para_9}) with triggering condition (\ref{trigger_3}) is adopted.

The parameters and functions in (\ref{follower_k}) and (\ref{condition1}) are given by $M=1650 (\text{kg})$, $\phi(v_f)=-0.1\big[1,v_f,v_f^2\big]^T$, $\theta=\left[0.5,5,0.25\right]^T$, $k_d=1.8$.
%The desired velocity of the following vehicle is set to be $v_d=20 (\text{m/s})$.
Suppose that the initial states are given by $x_l(0)=100 (\text{m})$, $v_l(0)=18 (\text{m/s})$, $x_f(0)=0 (\text{m})$, $v_f(0)=0 (\text{m/s})$,  $\hat{\theta}(0)=\left[0.4,4,0.3\right]^T$.
Control design parameters are chosen as $\gamma_0=0.2$, $\gamma_1=0.5$, $\Delta \tau_{\max}=8{\rm s}$, $k_1=1$, $k_2=2$, $c_1=0.25$, $c_2=0.25$, $H(x)={\rm diag}(1,10)$, $F(x)=\big[M \phi(v_f)^T \big(\hat{\theta}_V(t)+\hat{\theta}_h(t)\big),0\big]^T$ in Case I to Case III, $F(x)=\big[2M \phi(v_f)^T \hat{\theta}(t),0\big]^T$ in Case IV.

\begin{figure}[!htbp]
	\vspace{-0.2cm}
	\centering
	\includegraphics[width=8.2cm]{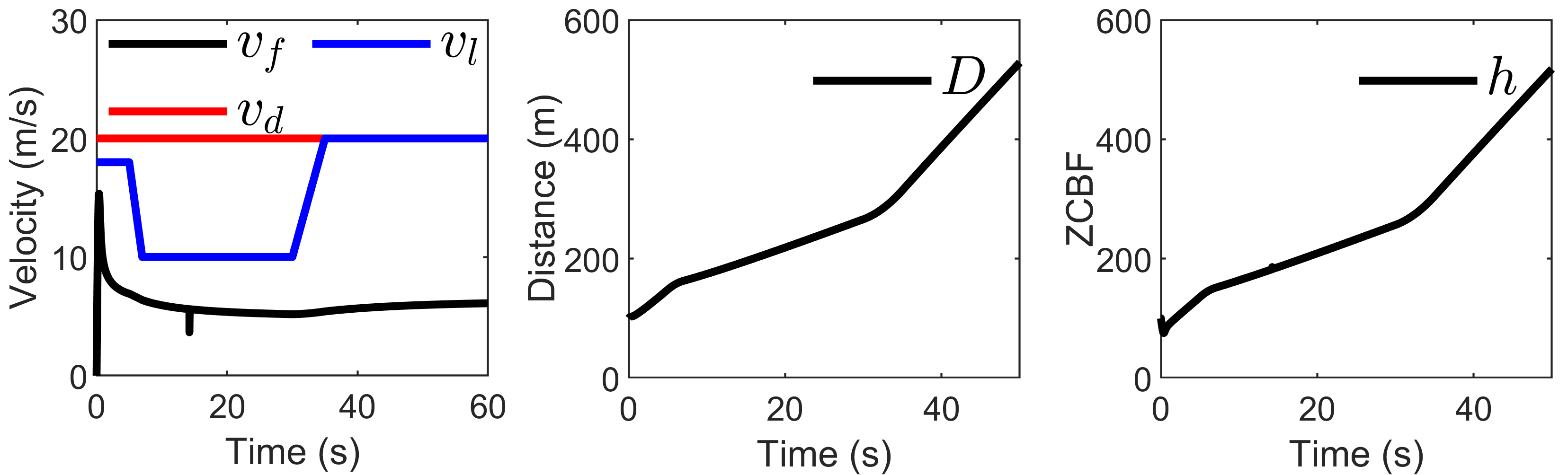}
	\vspace{-0.2cm}
	\caption{Velocity, distance and CBF in Case I.} 
	\label{Simu_Lyqp_1}
	\vspace{-0.2cm}
\end{figure}

\begin{figure}[!htbp]
	\vspace{-0.2cm}
	\centering
	\includegraphics[width=8.2cm]{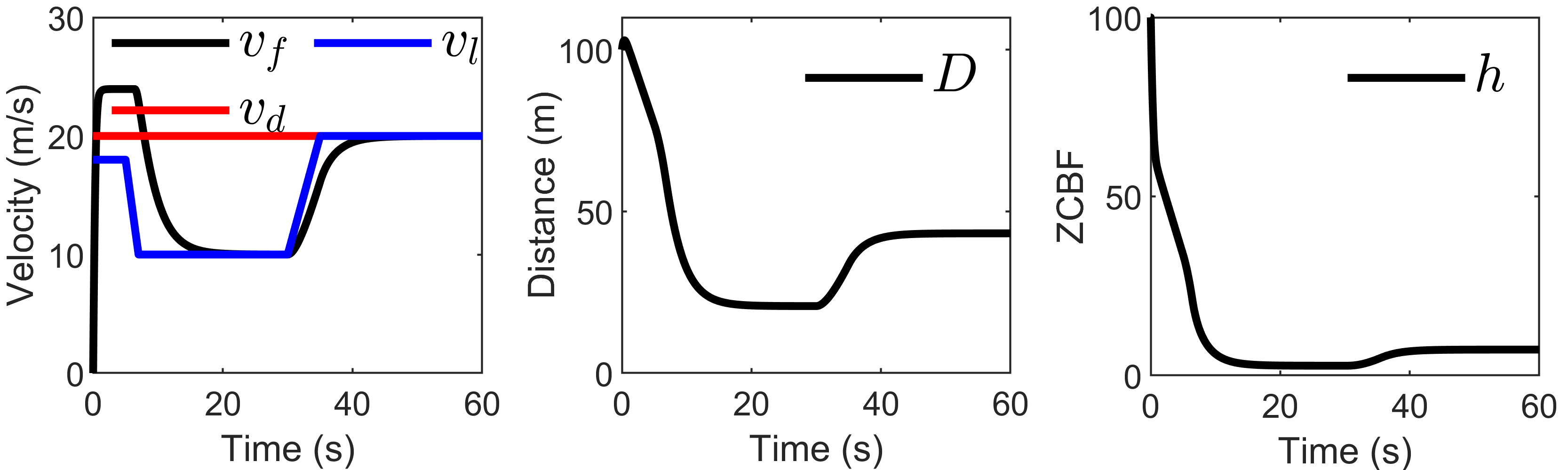}
	\vspace{-0.2cm}
	\caption{Velocity, distance and CBF in Case II.} 
	\label{Simu_Proj1_1}
	%\vspace{-0.2cm}
\end{figure}

\begin{figure}[!htbp]
	%\vspace{-0.2cm}
	\centering
	\includegraphics[width=8.2cm]{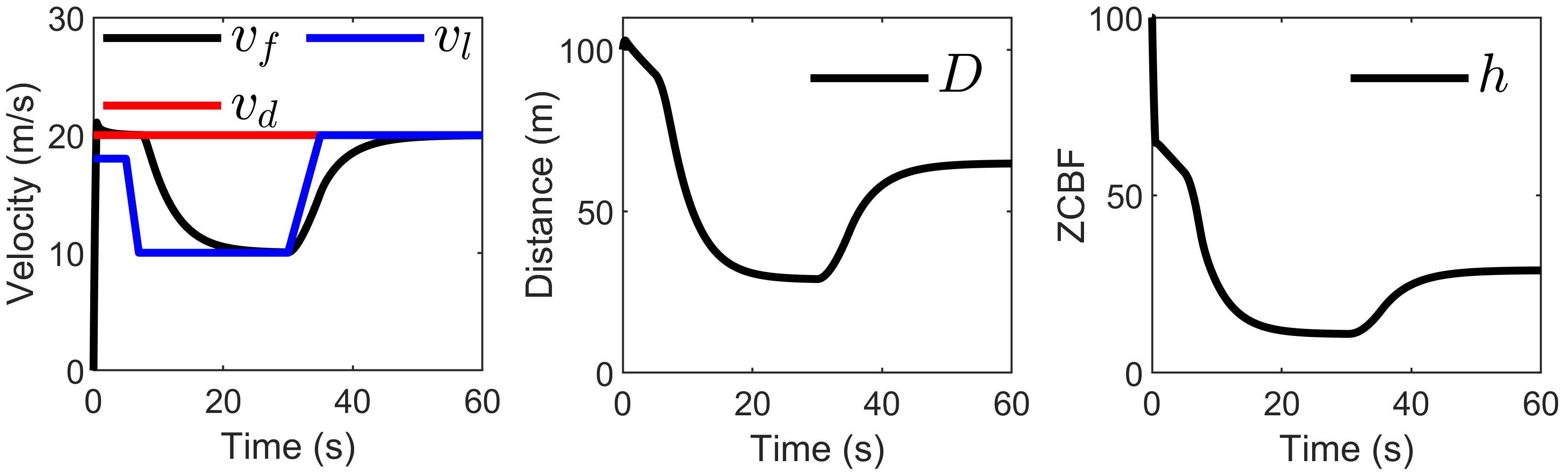}
	\vspace{-0.3cm}
	\caption{Velocity, distance and CBF in Case III.} 
	\label{Simu_Proj2_1}
	\vspace{-0.2cm}
\end{figure}

\begin{figure}[!htbp]
	%\vspace{-0.2cm}
	\centering
	\includegraphics[width=8.2cm]{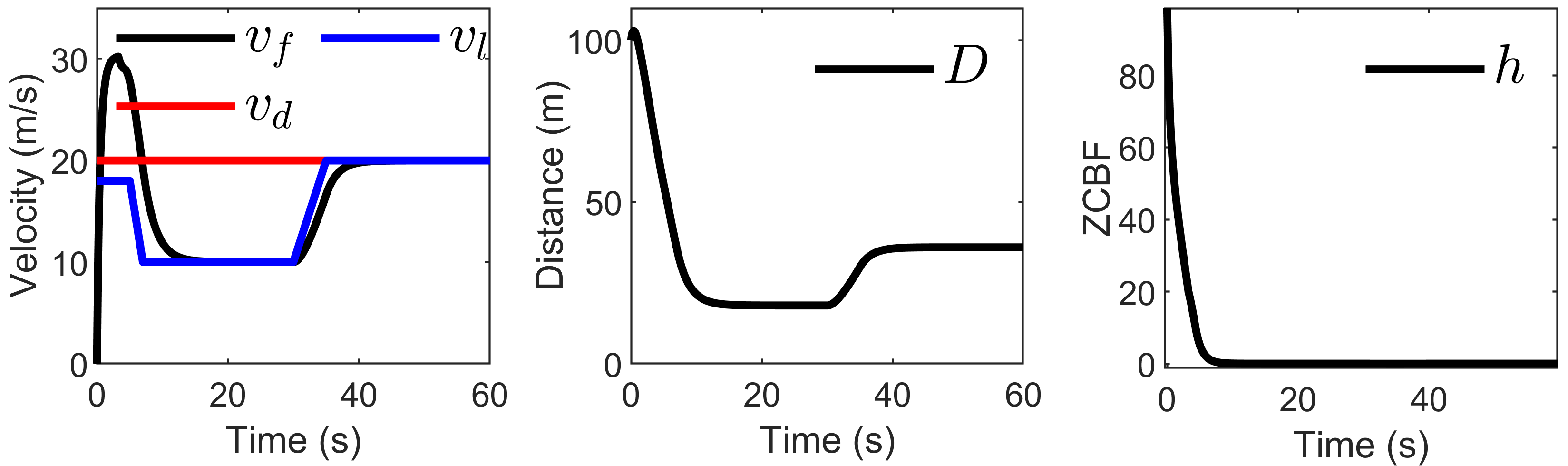}
	\vspace{-0.3cm}
	\caption{Velocity, distance and CBF in Case IV.} 
	\label{Simu_Event_1}
	\vspace{-0.2cm}
\end{figure}

\begin{figure}[!htbp]
	%\vspace{-0.2cm}
	\centering
	\includegraphics[width=8.2cm]{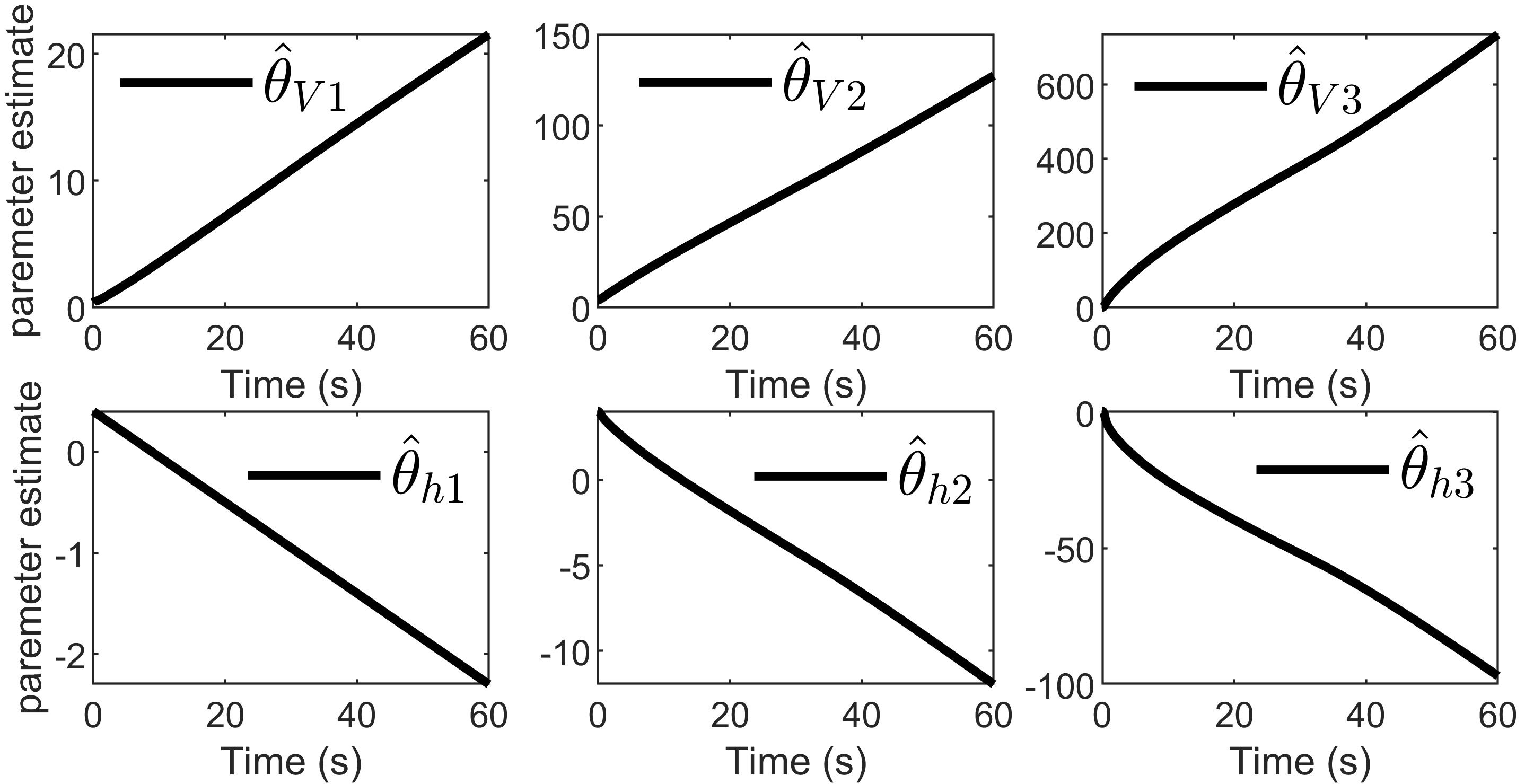}
	\vspace{-0.2cm}
	\caption{Parameter estimates in Case I.} 
	\label{Simu_Lyqp_2}
	\vspace{-0.2cm}
\end{figure}

\begin{figure}[!htbp]
	%\vspace{-0.2cm}
	\centering
	\includegraphics[width=8.2cm]{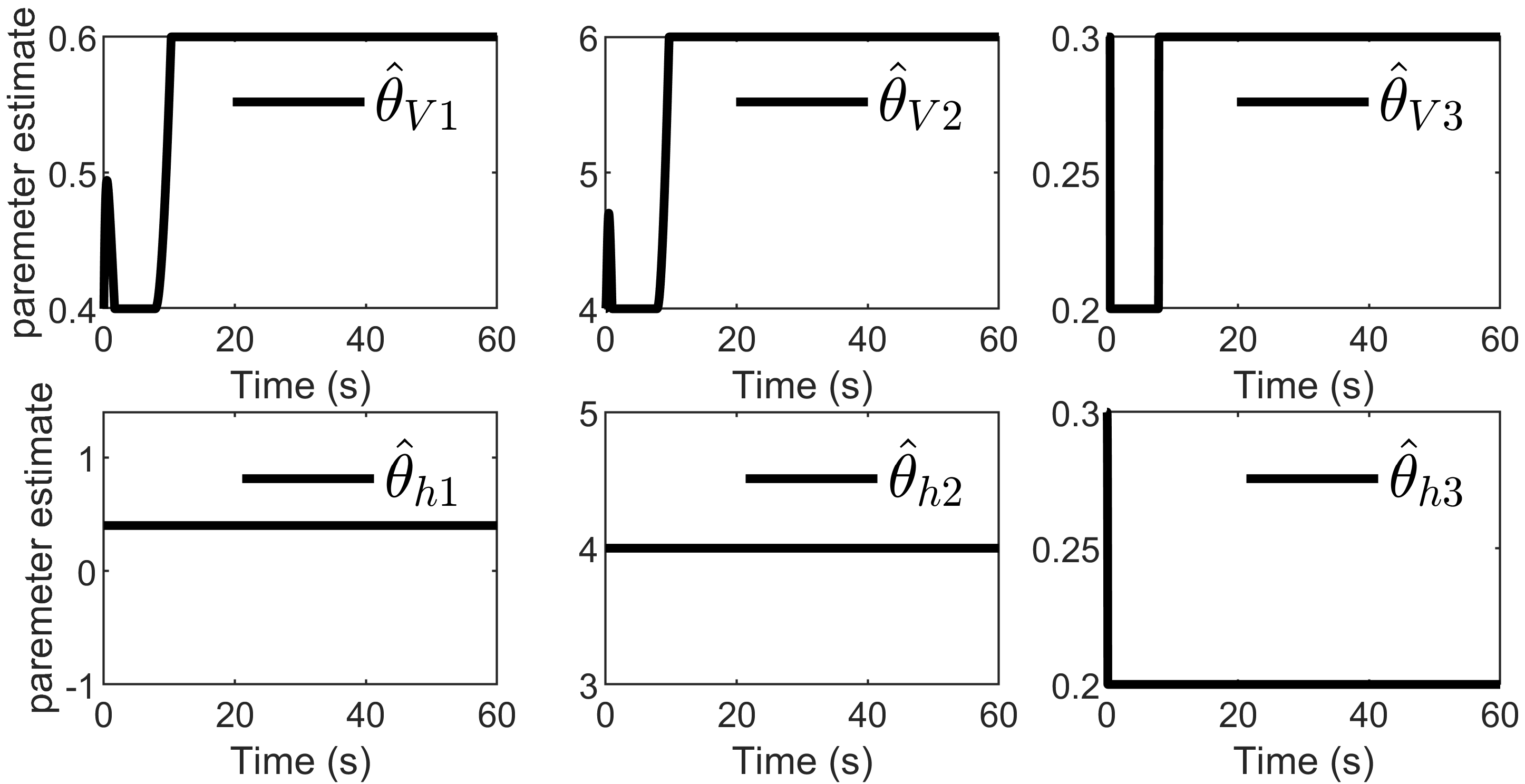}
	\vspace{-0.3cm}
	\caption{Parameter estimates in Case II.} 
	\label{Simu_Proj1_2}
	\vspace{-0.2cm}
\end{figure}

\begin{figure}[!htbp]
	%\vspace{-0.2cm}
	\centering
	\includegraphics[width=8.2cm]{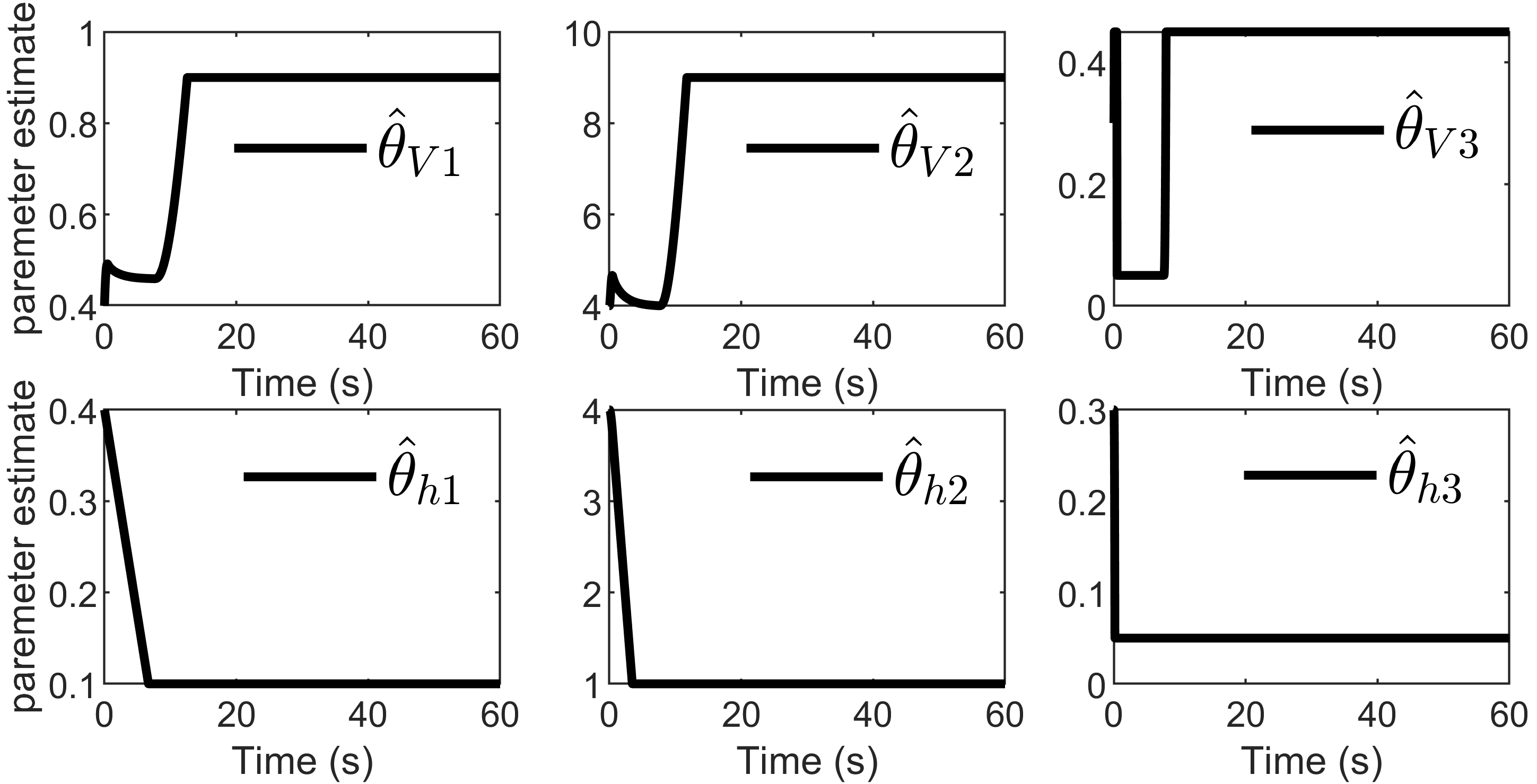}
	\vspace{-0.3cm}
	\caption{Parameter estimates in Case III.} 
	\label{Simu_Proj2_2}
	\vspace{-0.2cm}
\end{figure}

\begin{figure}[!htbp]
	%\vspace{-0.2cm}
	\centering
	\includegraphics[width=8.2cm]{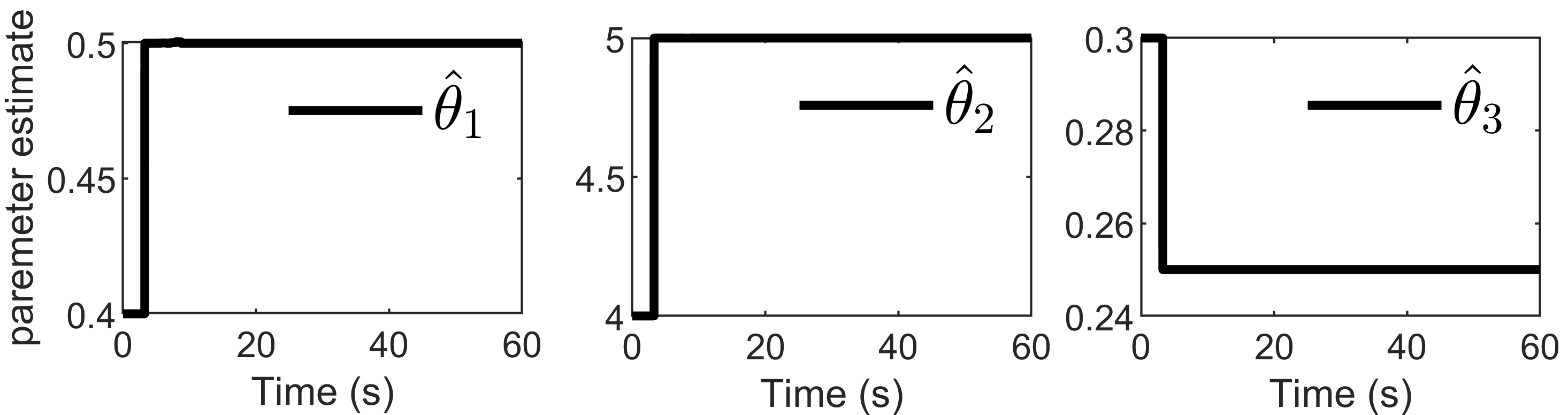}
	\vspace{-0.3cm}
	\caption{Parameter estimates in Case IV.} 
	\label{Simu_Event_2}
	%\vspace{-0.2cm}
\end{figure}

Fig. \ref{Simu_Lyqp_1} -- Fig. \ref{Simu_Event_1} show the velocities, distances and CBFs for Cases I -- IV, respectively.
Fig. \ref{Simu_Lyqp_2} -- Fig. \ref{Simu_Event_2} show the parameter estimates for Cases I -- IV, respectively.

It can be seen from Fig. \ref{Simu_Lyqp_2} that the amplitudes of certain parameter estimates in \textbf{Case I} keep increasing as time progresses, which supports the analysis in Remark 7.
Correspondingly in Fig. \ref{Simu_Lyqp_1}, the control performance is conservative, in the sense that the following vehicle keeps a long distance to the forward vehicle hence the speed tracking objective is totally abandoned. 

In Fig. \ref{Simu_Proj1_2} and Fig. \ref{Simu_Proj2_2}, the parameter estimates in \textbf{Case II} and \textbf{Case III} are constrained within the pre-known range by the projection operators.
However, as shown in Fig. \ref{Simu_Proj1_1} and Fig. \ref{Simu_Proj2_1}, the control performances of \textbf{Case II} and \textbf{Case III} are also conservative with inaccurate estimates.
Moreover, the control performance of \textbf{Case III} is worse than \textbf{Case II}, since the pre-known bounding set $\Theta_2$ is less accurate than $\Theta_1$. 
With the fact that the adaptive laws (\ref{Lyap_1}) and (\ref{Lyap_2}) are unbounded, the projection operators only guarantee boundedness of the estimates rather than providing an accurate estimate.
The control performance is tightly dependent on the accuracy of the pre-known set $\Theta$, which is similar to the ``worst-case uncertainty'' schemes, as discussed in Remark 8.

It is seen from Fig. \ref{Simu_Event_1} and Fig. \ref{Simu_Event_2} that the vehicle in \textbf{Case IV} attempts to track the desired velocity without violating the safety constraint.
The velocity tracking error and the tendency of $h(x)$ to 0 is inevitable, due to the potential conflict.
The simulation results in \textbf{Case IV} are satisfactory with accurate estimate and desired control performance, which demonstrates the effectiveness of the proposed adaptive control scheme.

\section{Conclusion}
In this paper, a triggered BaLSI-based adaptive safety scheme is proposed for uncertain systems with potentially conflicting control objectives and safety constraints.
The updating times of parameter estimates is no more than the dimension of the unknown parameter vector, and the effects due to unknown system parameters will be totally eliminated after the last updating moment of the parameter estimate.
A safety-triggered condition is presented, based on which  the forward invariant property of the safe set is guaranteed and the Zeno behavior can be excluded.
The adaptive law is designed by processing the data of system inputs and outputs, to avoid unbounded estimates and overparameterization problems in the existing results.
Future research efforts will be devoted towards considering delay analysis and ACC smooth transition.

\bibliographystyle{elsarticle-harv}      
\bibliography{reference}

\end{document}